\documentclass{article}
\usepackage{graphicx} 
\usepackage[english]{babel}
\usepackage[bottom=3.5cm, top=3.5cm]{geometry}
\usepackage{eurosym}
\usepackage{setspace}
\usepackage{appendix}
\usepackage{placeins} 
\usepackage{todonotes}
\usepackage{natbib}
\usepackage{bm}

\usepackage{fullpage}
\usepackage{xcolor}
\usepackage{hyperref}
\usepackage{comment}
\usepackage [autostyle, english = american]{csquotes} 
\usepackage{setspace}

\MakeOuterQuote{"} 

\graphicspath{ {./figs} }
\title{Structuring the Environment Nudges Participants Toward Hierarchical Over Shortest Path Planning}

\date{}

\author{Valeria Simonelli$^{*a,b}$, Davide Nuzzi$^{*a}$, Gian Luca Lancia$^a$, Giovanni Pezzulo$^{\dag a}$  \\
\\
$^a$ Institute of Cognitive Sciences and Technologies, National Research Council, Rome, Italy  \\
$^b$ University of Rome “La Sapienza”, Rome, Italy}

\footnotetext{These authors contributed equally to this work}
\footnotetext{Corresponding author at: Via Gian Domenico Romagnosi, 18, 00196 Roma (RM),Italia. E-mail address: giovanni.pezzulo@istc.cnr.it}

\begin{document}
\maketitle

\begin{abstract}

Effective planning is crucial for navigating complex environments and achieving goals efficiently. In this study, we investigated how environmental structure influences the selection of planning strategies. Forty-two participants navigated a space station to collect colored spheres, with environments either structured (spheres grouped by color) or unstructured (spheres scattered randomly). We tested three types of plans: hierarchical (grouping spheres by color), shortest path (minimizing travel distance), and neutral (none of the above). By manipulating environmental structure, we were able to nudge participants toward a preference for hierarchical planning in structured environments, while shortest path plans were favored in unstructured environments. A mismatch between self-reported preferences and actual choices indicated that participants often adopted implicit strategies, unaware of their decision-making processes. These findings highlight the powerful effect of environmental cues on planning and suggest that even subtle changes in structure can guide the selection of planning strategies.

\end{abstract}


\newpage
\section{Introduction}

Planning is a fundamental cognitive ability that allows individuals to organize and sequence actions to achieve a goal \citep{russell2016artificial,newell1958elements,mattar2022planning,dolan2013goals,hills2015exploration,parr2022active,hunt2021formalizing}. In daily life, planning occurs across a wide range of tasks, encompassing spatial navigation, object manipulation, and solving challenging puzzles \citep{eluchans2025adaptive,eluchans2024eye,maselli2023beyond,lancia2024separable,ho2022planning,fernandez2025expert,keramati2016adaptive,bongiorno2021vector,huys2015interplay,ribas2011neural,brunec2023exploration, simonelli2024human, nuzzi2024planning}. 

Previous studies have shown that, given our limited cognitive resources, individuals aim to optimize their use by balancing efficiency and cognitive effort \citep{lieder2020resource,simon1990bounded,bhui2021resource,callaway2022rational,gershman2015computational,lancia2023humans,genewein2015bounded}. However, there are various ways to balance efficiency and cognitive effort during planning. Take the example of planning a shopping trip to multiple stores in different parts of a city and deciding the order in which to visit them. One option is to visit stores randomly, without a structured plan. This random strategy would plausibly save cognitive resources at the expense of efficiency. Another possibility is to minimize travel distance by selecting the shortest path. This shortest path strategy is efficient but requires engaging cognitive effort to form a plan \citep{lancia2023humans,piray2021linear,todorov2009efficient,parr2023cognitive}. Alternatively, a hierarchical strategy involves grouping stores by location, visiting all stores in one neighborhood before moving to another. This approach divides the task into high-level (neighborhood) and low-level (stores within the neighborhood) components, potentially bypassing exhaustive cognitive search of a solution. However, it requires identifying the hierarchical structure to be exploited (e.g., the neighborhoods) \citep{botvinick2008hierarchical,botvinick2009hierarchically,donnarumma2016problem,solway2014optimal,tomov2020discovery,balaguer2016neural,pezzulo2018hierarchical,momennejad2020learning}. 

This raises an important question: What determines the trade-off between a shortest path strategy that minimizes travel distance (efficiency over cognitive effort) and a hierarchical planning strategy that uses the environmental structure to reduce the cognitive effort (cognitive effort over efficiency)? In this study, we aim to investigate the factors influencing participants' selection of planning strategies when navigating an environment to collect objects. We aim to test the hypothesis that participants' choice of planning strategy can be influenced by nudging an appropriate environmental structure. Specifically, we hypothesize that participants would prioritize the shortest path strategy in tasks without apparent structure and the hierarchical planning strategy in tasks in which we made environmental structure more salient. 

To address this hypothesis, we tested participants during a video game-like planning task, requiring them to navigate an astronaut from a third-person perspective through a space station to collect three or four colored spheres. The task allowed for three types of plans: \emph{shortest path} (i.e., collect the spheres by following the shortest path), \emph{hierarchical} (i.e., to group the spheres to be collected according to a higher-level organization -- color in our task -- and collect first all the spheres having the same color), or \emph{neutral} (i.e., none of the above). Crucially, the space station was either \emph{structured}, with spheres grouped by color, or \emph{unstructured}, with spheres scattered randomly. By analyzing participants' choices across different environments, we aimed to determine whether participants prioritized hierarchical versus shortest path planning in the presence or absence of environmental structure.

\section{Methods}

\subsection{Participants}

Forty-two participants (21 female, 20 male, 1 prefer not to say; \textcolor{black}{Mean age} = 30.8, SD = 7.55) from the Prolific platform took part in the study. Two were excluded for dropout, leaving 40 for analysis. Participants, aged 19-45, were from the USA or UK and were compensated for their 28-minute participation. All the procedures were approved by the CNR Ethics committee. 

We determined the sample size based on results from a pilot experiment, specifically the medians and standard deviations of participants' hierarchical choices in structured versus unstructured environments. Our calculations indicated that the selected sample size (40 participants) provides a confidence level (the minimum acceptable probability of preventing a Type I error) of $99.9\%$ and a test power (the minimum acceptable probability of preventing a Type II error) of $99\%$ for the Mann-Whitney test.

{\color{black}
\subsection{Data analysis methods}
We used the Mann–Whitney U test for all between-condition comparisons. This choice was based on visual inspection of the data, which revealed deviations from normality including skewness and outliers, consistent with prior recommendations \citep{micceri1989unicorn, erceg2008modern, rasch2011two, ghasemi2012normality}. When more than one comparison was performed within a family of tests, we applied the Benjamini–Hochberg False Discovery Rate correction \citep{benjamini1995controlling}. Effect sizes are reported as $r$, derived from the U-statistic of the Mann–Whitney test and Wendt formula \citep{wendt1972dealing}.
}



\subsection{Experimental task}

\begin{figure}[h!]
    \centering
    \includegraphics[width=1\linewidth]{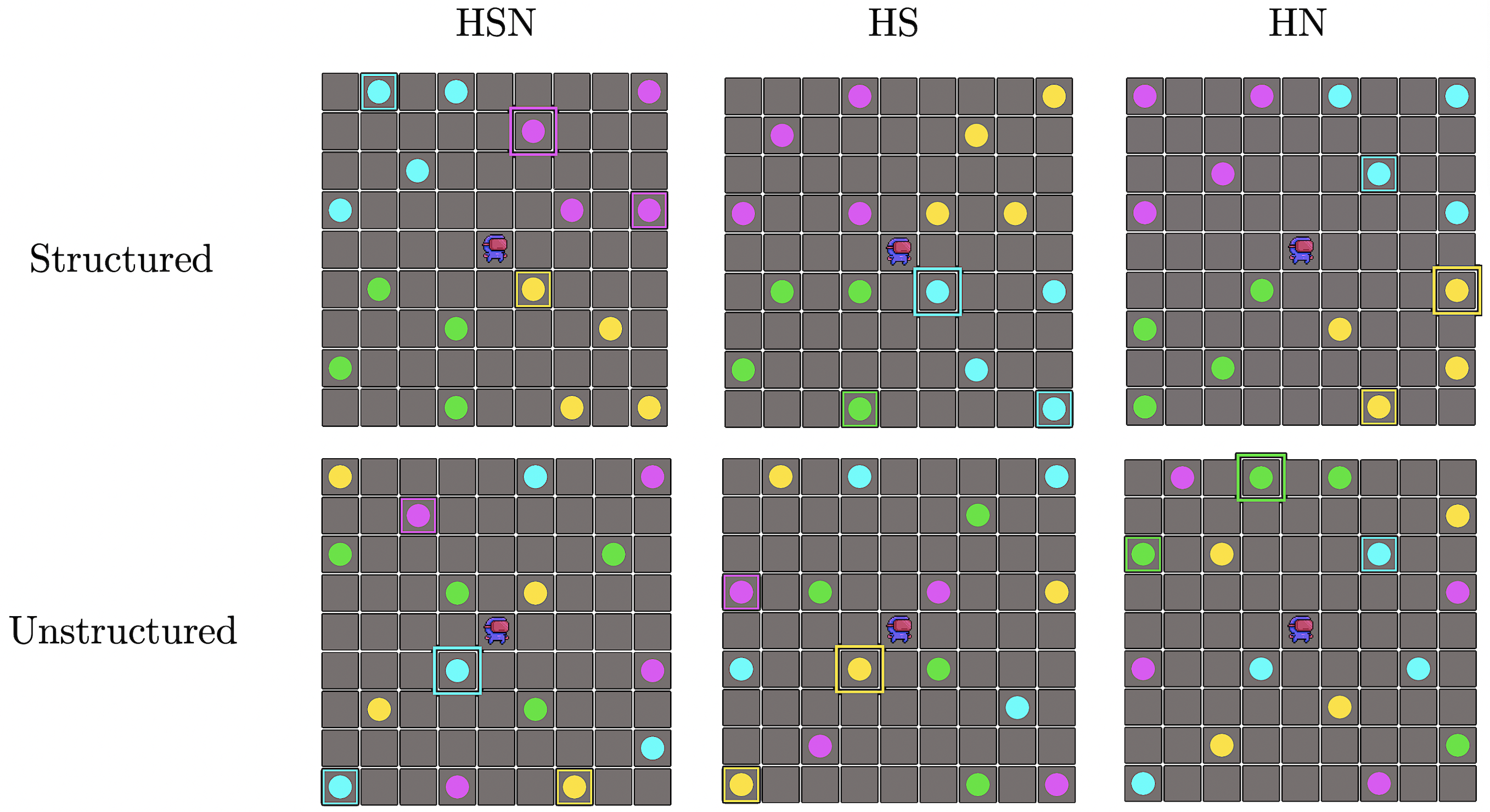}
    \caption{The space station game. The figure shows six example trials, representing all possible combinations of environmental structure (either "structured", where spheres are grouped by color, or "unstructured", where spheres are not grouped) and plan availability (either "HSN trials", with hierarchical, shortest, and neutral plans; "HS trials", with hierarchical and shortest plans; or "HN trials", with hierarchical and neutral plans available). At the beginning of each trial, the astronaut is located at the center of the space station, with a 30-second countdown visible in the top left corner of the screen and a trial counter in the top right corner. Each space station contains 16 colored spheres and, among them, three or four are marked with colored frames. During the trial, participants can navigate the space station to reach and collect the spheres by pressing the space bar on their keyboard. Successfully completing a trial requires first collecting the sphere marked with the largest frame ("forced choice"), followed by the other spheres marked with smaller frames in any order the participants prefer ("free choices"). The trial ends either when all framed spheres are collected or if time expires before all the framed spheres are collected. The time limit is not strict, as participants typically complete trials in approximately nine seconds on average. The image shown here depicts a slightly modified scenario compared to what participants experienced, in order to enhance visibility due to the small dimensions of the figure. See the main text for explanation.}
    \label{fig:experimental_setup}
\end{figure}

Participants played a computer-based game created with Unity (Unity Technologies, San Francisco, US). They controlled an avatar ("astronaut") tasked with navigating a $9\times9$ grid space ("space station") to collect three or four colored spheres (Figure \ref{fig:experimental_setup}). Participants moved the avatar by using the WASD or arrow keys (up, down, left, right) and collected the spheres using the space bar of their keyboard. 

The space station contained 16 colored spheres, arranged in two ways across 2 experimental blocks of 60 trials each (counterbalanced across participants). In the "structured" block, the spheres were grouped by color, whereas in the "unstructured" block, they were not. For each trial, three or four spheres were marked with colored frames, indicating that they had to be collected. The larger frame indicated that the sphere was the first to be collected (i.e., forced choice), while the smaller frames indicated the other spheres, which participants could collect in whatever order they preferred (i.e., free choices). Crucially, we designed trials in such a way that participants' first free choice would reveal their preference for one of the following possible plans: a plan that starts with the sphere having the same color as the one collected with the forced choice (\emph{hierarchical plan}), a plan that starts with the closest sphere (\emph{shortest plan}), or a plan that starts with neither of them (\emph{neutral plan}). \textcolor{black}{Note that we refer to the first type of plan as "hierarchical" because it organizes the spheres spatially according to a higher-order principle—in this case, color—such that spheres of the same color are located near one another. This is analogous to grouping stores by neighborhood in the shopping example introduced earlier.} Importantly, both the hierarchical and neutral plans required the same number of steps to collect the first "free choice" sphere, while the shortest path required one less step.

Participants first familiarized themselves with the game by completing four practice trials, which were excluded from the analysis. The experiment comprised two blocks, one \emph{structured} and one \emph{unstructured}, counterbalanced across participants. Each block consisted of 60 trials. To introduce variety and prevent participants from developing overly rigid strategies, we subdivided each block into three sets of 20 trials: \emph{HN trials}, \emph{HS trials}, and \emph{HSN trials}, each offering different plan options. In the \emph{HN trials}, participants had to collect three spheres, with hierarchical and neutral plans available. In the \emph{HS trials}, participants had to collect three spheres, with hierarchical and shortest plans available. Finally, in the \emph{HSN trials}, participants had to collect four spheres, with all three plans—hierarchical, shortest, and neutral—simultaneously available. The order of presentation of these 60 trials within each block was randomized for each participant.


This design allows us to investigate the effects of environmental structure ("structured" vs. "unstructured" blocks) on participants' choice of planning strategies. We were primarily interested in participants' choice between hierarchical plans that simplify decision-making versus shortest path plans that provide more efficient solutions. We hypothesized that participants' choices would depend on the saliency of environmental structure: in "structured" blocks, in which spheres are organized by color, participants would prefer hierarchical plans -- first collecting all spheres of the same color -- reflecting their sensitivity to environmental structure; whereas in "unstructured" blocks, participants would prefer shortest plans to complete the task faster. Additionally, we were interested in the choice between hierarchical versus neutral plans, which have the same length. We wanted to test whether, regardless of the saliency of the environment, participants would prefer the hierarchical plan, reflecting an implicit propensity to follow strategies that simplify decision-making where there is nothing to earn by selecting the opposite (neutral) plan. Finally, we were interested in testing whether participants' reaction time and movement time during the experiment were influenced by their choice among hierarchical, shortest, and neutral plans.

After the experiment, we asked participants to self-report their preference for the planning strategies by posing the following question: 'On a scale from 1 to 5, with 1 indicating a strong influence from distance (shortest plans) and 5 indicating a strong influence from color (hierarchical plans), how much do you think your choices were influenced by these factors? A response of 3 indicates no influence from either.' We reasoned that high versus low levels of congruence between participants' choices and self-reports might indicate the use of more explicit versus more implicit planning strategies.


\section{Results}

\subsection{Choice between hierarchical, shortest and neutral plans}

\begin{figure}[h!]
    \centering
    \includegraphics[width=0.45\linewidth]{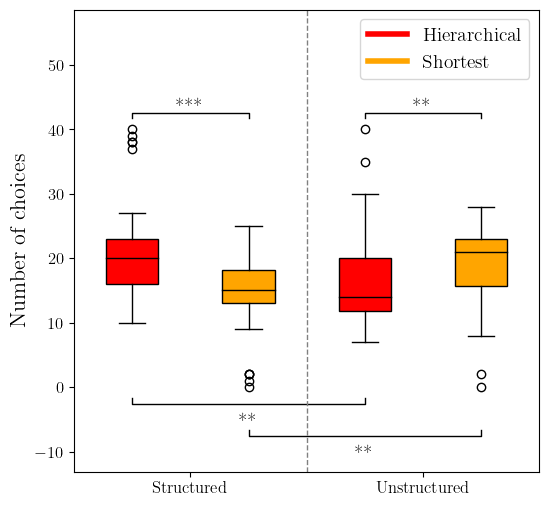}
    \includegraphics[width=0.45\linewidth]{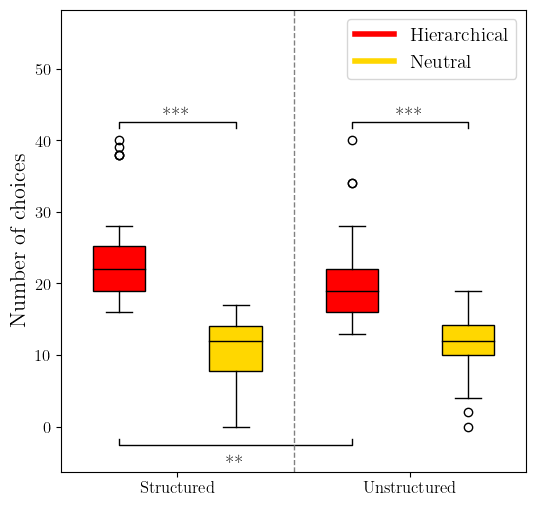}
    \caption{\textbf{Number of choices between plans.} Left: Boxplots showing the number of choices made by participants in structured vs. unstructured environments, considering hierarchical versus shortest path plans (aggregating all data from HS e HSN trials). Right: Boxplots showing the number of choices made by participants in structured vs. unstructured environments, considering hierarchical versus neutral plans (aggregating all data from HN e HSN trials). Level of significance is assessed via the Mann-Whitney test and the FDR correction is applied. Significant differences between conditions are indicated (*$p < 0.05$, **$p < 0.01$, ***$p < 0.001$).}
    \label{fig:Results}
\end{figure}

To test participants' preference for hierarchical, shortest or neutral plans in structured and unstructured blocks, we performed Mann-Whitney tests with False Discovery Rate (FDR) correction. We made two analyses, the former aggregating all experimental trials (HS and HSN) in which hierarchical and shortest path plans were available, and the latter aggregating all experimental trials (HN and HSN) in which hierarchical and neutral plans were available (Figure \ref{fig:Results}). 

We fist considered all trials in which participants selected between hierarchical and shortest plans, in both structured and unstructured blocks (Figure \ref{fig:Results}, left). For this, we aggregated data from HS and HSN trials. Participants selected significantly more hierarchical plans compared to shortest plans in structured blocks ($p< 0.001$, \textcolor{black}{$r = 0.50$}). In contrast, in unstructured blocks, participants selected significantly more shortest plans compared to hierarchical plans ($p < 0.01$, \textcolor{black}{$r = -0.33$}). Furthermore, participants selected more hierarchical plans in structured compared to unstructured blocks ($p < 0.01$, \textcolor{black}{$r = 0.45$}) and more shortest plans in unstructured compared to structured blocks ($p < 0.01$, \textcolor{black}{$r = -0.44$}). These results highlight that hierarchical planning dominates in structured environments, while shortest path planning dominates in unstructured environments. Furthermore, the presence of structure in the environment increases the preference for hierarchical planning, whereas its absence leads to a greater preference for shortest path planning.

We next considered all trials in which participants selected between hierarchical and neutral plans, in both structured and unstructured blocks (Figure \ref{fig:Results}, right). For this, we aggregated data from HN and HSN trials. Participants selected significantly more hierarchical plans compared to neutral plans in both structured ($p < 0.001$, \textcolor{black}{$r = 1$}) and unstructured blocks ($p < 0.001$, \textcolor{black}{$r = 0.79$}). Furthermore, participants selected more hierarchical plans in structured compared to unstructured blocks ($p < 0.01$, \textcolor{black}{$r = 0.38$}). These results highlight the dominance of hierarchical planning when the alternative (neutral) plan has the same path length and that environmental structure increases the dominance of hierarchical over neutral planning. Separate analyses of HN, HS, and HSN trials reveal the same qualitative pattern of results (Supplementary Section \ref{sec:separate}).

{\color{black}
We performed additional analyses comparing participants who began with the structured versus unstructured block, in order to test for possible order effects on plan choices. Specifically, we compared the number of choices made by participants who started with the structured block versus those who started with the unstructured block. These comparisons were conducted separately for each trial type (HN, HS, HSN), environment structure (structured, unstructured) and choice type (hierarchical, shortest, neutral), using Mann–Whitney U tests. We also repeated the analysis after aggregating across trial types. None of these comparisons yielded significant differences.

}


\subsection{Movement times}

To assess participants' movement times during the experiment, we used Mann-Whitney tests with False Discovery Rate (FDR) correction. For consistency, we grouped the data in the same way as in the analysis of participants' choices between plans. We analyzed both the (z-scored) movement time to collect the first sphere (forced choice) and the second sphere (free choice). The movement time for the first sphere was defined as the duration from trial onset to its collection, while the movement time for the second sphere was defined as the duration between collecting the first and second spheres.

\begin{figure}[h!]
    \centering
    \includegraphics[width=\linewidth]{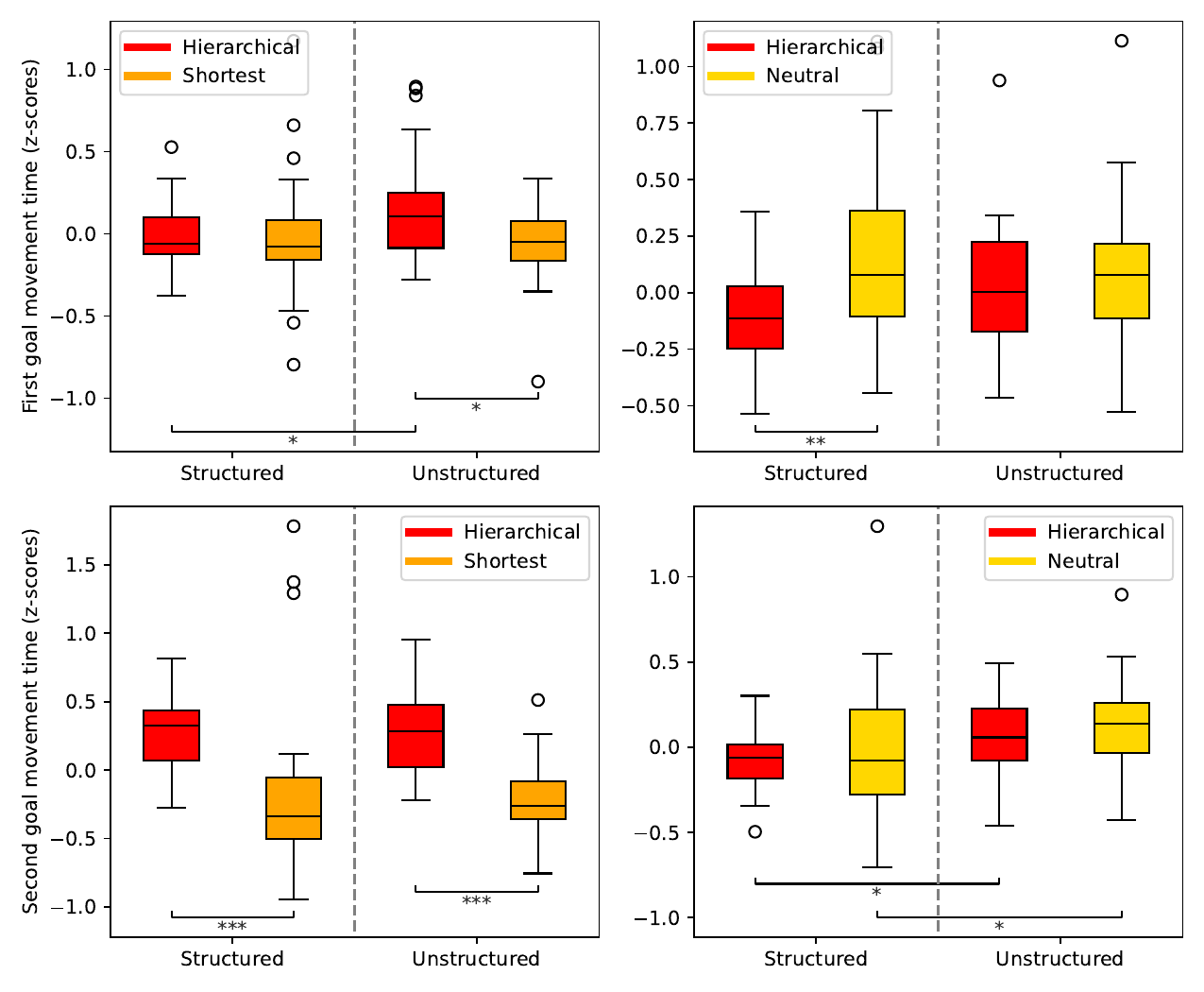}
\caption{\textbf{Standardized movement times (z-scores) per subject.} Top row: Boxplots of movement times to the first sphere (forced choice), defined as the duration from trial onset until collection of the first sphere. The left panels show hierarchical versus shortest-path plans (aggregating data from HS and HSN trials) and the right panels show hierarchical versus neutral plans (aggregating data from HN and HSN trials). Bottom row: Boxplots of movement times to the second sphere, defined as the duration between collecting the first sphere and collecting the second sphere, under the same planning conditions. Significance was assessed via the Mann–Whitney test with FDR correction (significant differences: *$p < 0.05$, **$p < 0.01$, ***$p < 0.001$).}
    \label{fig:MovementTime}
\end{figure}

We first considered all trials in which participants made the initial forced choice, selecting between the hierarchical and shortest plans (Figure \ref{fig:MovementTime}, top left) and between the hierarchical and neutral plans (Figure \ref{fig:MovementTime}, top right) in both structured and unstructured blocks. Participants collected the first sphere faster when selecting the shortest plan over the hierarchical plan in unstructured blocks ($p < 0.05$, \textcolor{black}{$r = 0.36$}) and when selecting the hierarchical plan over the neutral plan in structured blocks ($p < 0.01$, \textcolor{black}{$r = -0.44$}). Additionally, when collecting the first sphere and choosing the hierarchical plan over the shortest plan, participants were faster in structured compared to unstructured blocks ($p < 0.05$, \textcolor{black}{$r = -0.32$}).

We next considered all trials in which participants made the second, free choice, selecting between hierarchical and shortest plans (Figure \ref{fig:MovementTime}, bottom left) and between hierarchical and neutral plans (Figure \ref{fig:MovementTime}, bottom right) in both structured and unstructured blocks. Participants collected the second sphere faster when selecting the shortest plan over the hierarchical plan in both structured ($p < 0.001$, \textcolor{black}{$r = 0.73$}) and unstructured blocks ($p < 0.001$, \textcolor{black}{$r = 0.81$}). Additionally, when collecting the second sphere, they were faster in structured compared to unstructured blocks, both when selecting the hierarchical plan ($p < 0.05$, \textcolor{black}{$r = -0.39$}) and the neutral plan ($p < 0.05$, \textcolor{black}{$r = -0.30$}).

\subsection{Reaction times}

To assess participants' reaction times during the experiment, we used Mann-Whitney tests with False Discovery Rate (FDR) correction. For consistency, we grouped the data in the same way as in the analysis of participants' choices between plans. We analyzed reaction times for collecting both the first sphere (forced choice) and the second sphere (free choice). The reaction time for the first sphere was defined as the latency from trial onset to the initiation of the first movement (i.e., the first movement toward the first sphere). The reaction time for the second sphere was defined as the latency from the collection of the first sphere to the initiation of the movement toward the second sphere.

\begin{figure}[h!]
    \centering
    \includegraphics[width=\linewidth]{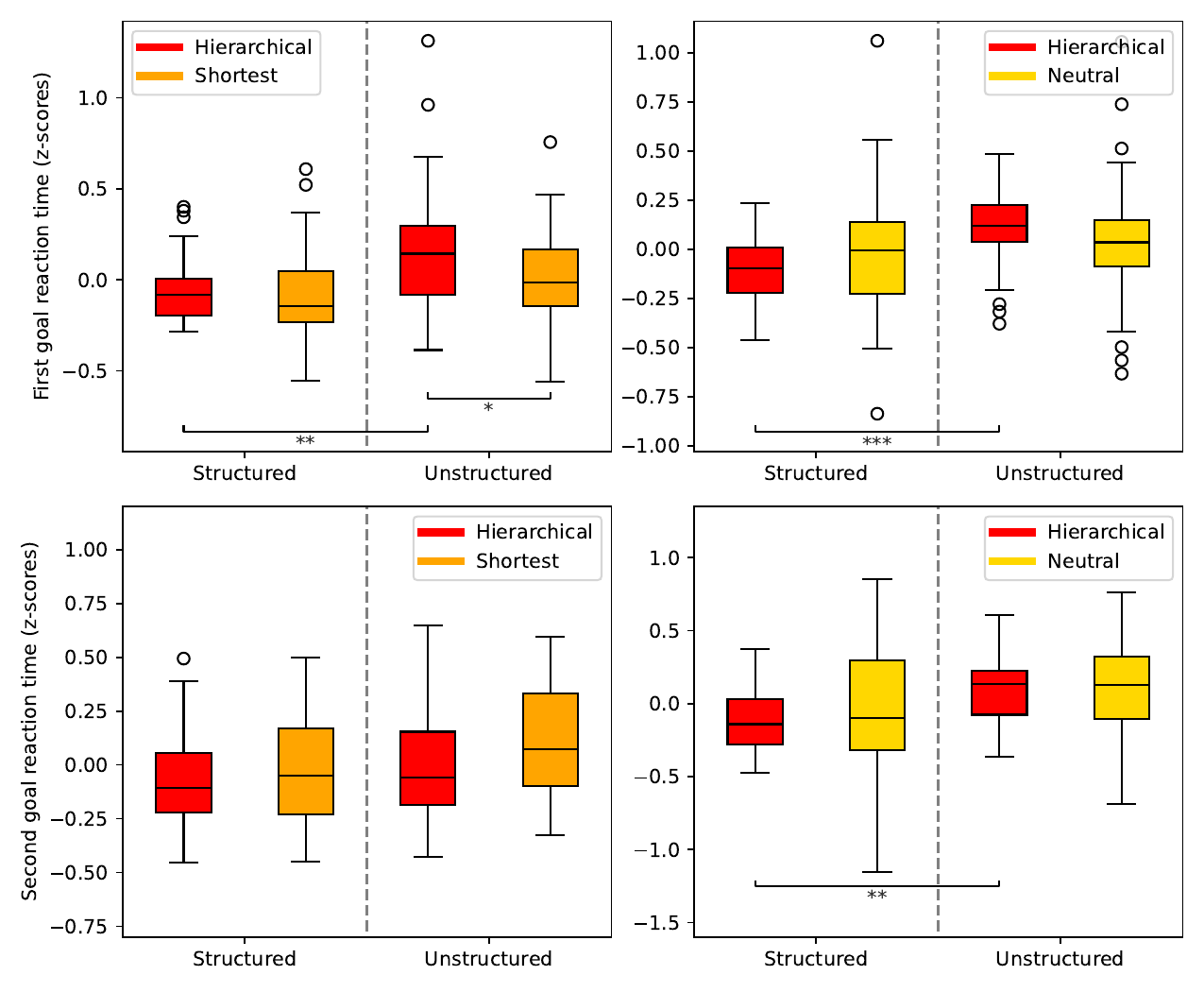} 
\caption{\textbf{Standardized reaction times (z-scores) per subject.} Top row: Boxplots of reaction times for the first sphere, defined as the latency from trial onset until initiation of the first movement (i.e., the first movement toward the first sphere), under two planning conditions: the left panels show hierarchical versus shortest-path plans (aggregating data from HS and HSN trials) and the right panels show hierarchical versus neutral plans (aggregating data from HN and HSN trials). Bottom row: Boxplots of reaction times for the second sphere, defined as the latency following collection of the first sphere until initiation of the movement toward the second sphere, under the same planning conditions. Significance was assessed via the Mann–Whitney test with FDR correction (significant differences: *$p < 0.05$, **$p < 0.01$, ***$p < 0.001$).}
    \label{fig:ReactionTime}
\end{figure}

We first considered all trials in which participants made the initial forced choice, selecting between the hierarchical and shortest plans (Figure \ref{fig:ReactionTime}, top left) and between the hierarchical and neutral plans (Figure \ref{fig:ReactionTime}, top right) in both structured and unstructured blocks. When moving toward the first sphere, participants had a faster reaction time when selecting the shortest plan over the hierarchical plan in unstructured blocks ($p < 0.05$, \textcolor{black}{$r = 0.30$}). Additionally, participants showed faster reaction times in structured compared to unstructured blocks, both when selecting the hierarchical plan over the shortest plan ($p < 0.01$, \textcolor{black}{$r = -0.47$}) and when selecting the hierarchical plan over the neutral plan ($p < 0.001$, \textcolor{black}{$r = -0.54$}).

We next considered all trials in which participants made the second, free choice, selecting between the hierarchical and shortest plans (Figure \ref{fig:ReactionTime}, bottom left) and between the hierarchical and neutral plans (Figure \ref{fig:ReactionTime}, bottom right) in both structured and unstructured blocks. When moving toward the second sphere and selecting the hierarchical plan over the neutral plan, participants had a faster reaction time in structured compared to unstructured blocks ($p < 0.01$, \textcolor{black}{$r = -0.46$}).

\subsection{Congruency between choices and self-reports}

\begin{figure}[h!]
    \centering
    \includegraphics[width=0.6\linewidth]{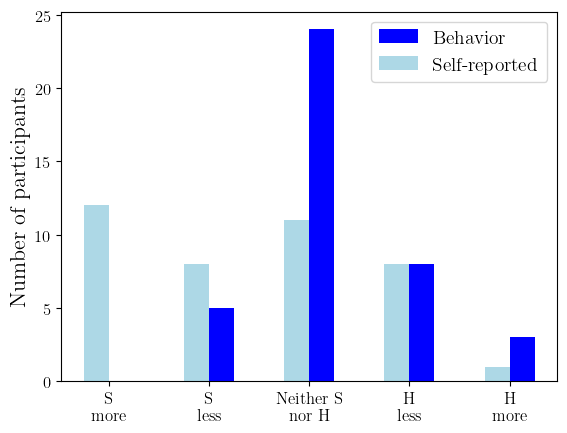}
    \caption{Bar plots comparing participants' behavior during the task (blue) and their self-reported choices (light blue).
    \textcolor{black}{Both distributions are mapped onto a shared 5-point scale: “S more” (strong preference for shortest plans), “S less” (moderate preference for shortest plans), “Neither S nor H” (no clear preference), “H less” (moderate preference for hierarchical plans), and “H more” (strong preference for hierarchical plans). Self-reported data reflect participants’ responses to a post-task questionnaire. Behavioral data are based on a normalized index of each participant’s choices, discretized to match the scale of the self-report. See Section 3.4 for full details on the computation of the behavioral index.}
    }

    \label{fig:Supplementary_behavior_vs_selfreported}
    \end{figure}

Finally, we examined whether participants were accurate in their self-reports of a preference for hierarchical or shortest plans.
{\color{black}
To allow for a direct comparison between behavioral and self-reported preferences, we computed a behavioral preference index for each participant. This index was defined as $k = (n_H - n_S)/(n_H + n_S)$, where $n_H$ and $n_S$ are the total number of hierarchical and shortest-path choices, respectively, made by the participant across all trials. The resulting value falls within the range $[-1, 1]$, where $-1$ and $-1$ indicates exclusive preference for shortest-path plans, and $+1$ exclusive preference for hierarchical plans. For comparison with the 5-point self-report scale, this continuous index was discretized into five equally spaced categories. These behavioral categories are derived from the same dataset reported in previous sections, but aggregated and rescaled to match the format of the self-report responses.}
Interestingly, we found that participants' self-report responses significantly differed from their actual choices of hierarchical or shortest plans ($p < 0.01$, \textcolor{black}{$r = 0.36$}), with participants inaccurately reporting a greater preference for shortest plans compared to their actual choices (Figure \ref{fig:Supplementary_behavior_vs_selfreported}). This result suggests that participants may primarily adopt implicit planning strategies without being consciously aware of them. More speculatively, it suggests that participants might mistakenly perceive hierarchical plans as the shortest.

\section{Discussion}

\begin{quote}
    "Human rational behaviour is shaped by a scissors whose blades are the structure of task environments and the computational capabilities of the actor." \citep{simon1990invariants}. 
\end{quote}


This study aimed to investigate the factors influencing participants' selection of planning strategies when navigating an environment to collect objects, specifically examining how environmental structure affects the preference for hierarchical versus shortest path planning. The results reveal clear patterns in how participants adapt their planning strategies based on the structure of the environment and how these strategies are reflected in their implicit choices.

The central finding of this study is that environmental structure plays a significant role in shaping participants’ planning strategy preferences. In structured environments, where spheres were grouped by color, participants overwhelmingly preferred hierarchical plans, which involve grouping goals according to a higher-level organization (e.g., color in our task). Interestingly, participants’ preference for hierarchical plans was observed even when it implied discarding the shortest path. This suggests that participants were not purely focused on task completion time but also on the cognitive simplicity of the task. Conversely, in unstructured environments where spheres were scattered randomly, participants were more likely to select shortest path plans, as these offered a more efficient solution to the task. Further supporting the differing prevalence of hierarchical and shortest path plans in structured versus unstructured environments, the use of hierarchical plans increased significantly in structured environments, while the use of shortest path plans increased significantly in unstructured environments.

This finding aligns with the notion that individuals often seek to balance cognitive effort and task efficiency \citep{lieder2020resource,simon1990bounded,bhui2021resource,callaway2022rational,gershman2015computational,lancia2023humans}. However, this study extends previous work by demonstrating that the trade-off between cognitive effort and task efficiency can be influenced by subtle changes in environmental structure. It also helps systematize previous findings showing that participants prioritize hierarchical versus shortest path (or more broadly, efficient) solutions in different conditions. For example, neighborhood structure during navigation and problem solving promotes hierarchical strategies \citep{solway2014optimal,donnarumma2016problem,balaguer2016neural,correa2025exploring,eckstein2020computational,eckstein2021mind,zhu2023intention}. Conversely, when the environment lacks clear structure or when efficiency is prioritized, participants tend to select efficient paths, though not always the shortest \citep{lancia2023humans,bongiorno2021vector}. While previous findings demonstrate that people prioritize different strategies depending on the context, direct comparisons are challenging due to the use of distinct experimental designs. \textcolor{black}{A key contribution of this study, compared to previous research, is the finding that even small changes in environmental structure—such as the spatial arrangement of colors—are sufficient to influence participants’ planning preferences. This suggests that switching between different planning strategies may be easier than commonly assumed, highlighting the flexibility and context-dependence of human planning behavior.}

Another finding of this study is that the preference for hierarchical over equally efficient (neutral) strategies is evident in both structured and unstructured environments. An advantage of hierarchical planning was expected, given that selecting neutral plans yielded no benefit. However, even when choosing between hierarchical and neutral plans, participants demonstrated sensitivity to environmental structure, as indicated by their increased preference for hierarchical plans in structured environments. This result confirms individuals' sensitivity to environmental structure, particularly when it is made more salient.

Our study also demonstrates that participants' movement time and reaction time are influenced by plan selection. The patterns observed in movement and reaction times show similarities to the choice results. Specifically, we find a relative advantage of the shortest plan over the hierarchical plan in unstructured environments, an advantage of the hierarchical plan over the neutral plan in structured environments, and overall faster performance with the hierarchical plan in structured compared to unstructured blocks. \textcolor{black}{Moreover, participants are faster in completing the trials in structured compared to unstructured environments.} Some of these findings are expected -- for example, the fact that participants reach the second goal more quickly when selecting the shortest path is intuitive, as it minimizes travel distance. However, the observation that differences in movement and reaction times emerge even during the initial, forced choice is more surprising. This suggests that participants might engage in early preplanning, determining the sequence of spheres they intend to collect before making their first free choice. Moreover, the faster reaction times when choosing hierarchical over neutral plans in structured blocks further support the idea that participants leverage structured environments to facilitate decision-making. \textcolor{black}{Additionally, the fact that participants complete the task more quickly in structured environments suggests that making structure more salient not only shifts their preference toward hierarchical plans, but also enhances overall task efficiency.}

Finally, another interesting finding of this study is the mismatch between participants’ self-reported preferences for hierarchical or shortest path plans and their actual choices. Participants tended to report a stronger preference for shortest plans than was reflected in their behavior. This discrepancy suggests that participants may not be fully aware of the strategies they adopt during the task, highlighting the implicit nature of their decision-making processes. It is also possible that participants misperceive hierarchical plans as the shortest, conflating cognitive efficiency with physical distance. This speculative hypothesis remains to be tested in future research. \textcolor{black}{This could be achieved, for example, by presenting participants with the same problems used in this task and explicitly asking them to select the shortest path or to identify which path is shortest, rather than allowing them to solve the problems freely, as was done in the current study.}

These results have important implications for our understanding of how individuals select planning strategies. They suggest that the preference for hierarchical versus shortest path planning can be flexibly influenced by introducing exploitable structure in the environment. When the structure is made more salient and easily identifiable, individuals tend to exploit it to simplify decision-making and reduce cognitive load. They also select hierarchical plans when the alternatives show no advantage. These findings can inform the design of more effective strategies to encourage hierarchical planning in fields like education, problem solving and urban design, by creating appropriately structured environments, aligning with the concept of affordance in ecological psychology \citep{gibson1966senses, pezzulo2016navigating, rietveld2017optimal}.

While this study provides valuable insights into the factors influencing planning strategy selection, it is important to consider its limitations. 

\textcolor{black}{First, the task took place in a simplified game setup, which we called "environment" for simplicity, but which lacks the realism of ecological environments. Furthermore, the task was relatively constrained in terms of the types of plans available (hierarchical, shortest, and neutral). These issues may limit the generalizability of the findings to more complex, real-world tasks. Future research could address how environmental cues influence the choice between multiple available path plans in ecologically valid environments, such as cities \citep{golledge1976spatial,garling1988distance,garling1993understanding,car1994general,bongiorno2021vector, griesbauer2025london, fernandez2025expert}.} \textcolor{black}{Moreover, future studies could more systematically investigate how participants balance cognitive effort and task efficiency during planning. One approach would be to vary parametrically the difference in length between the shortest and hierarchical plans, to determine the point at which participants stop preferring the hierarchical plan in structured environments. Another approach could involve testing alternative ways to bias participants toward one strategy over the other—for example, by constraining the time available to complete a trial (favoring efficiency and shorter paths) or introducing a cognitively demanding dual task (favoring less effortful planning strategies). Another intriguing question for future research is whether and how people combine shortest-path and hierarchical planning to navigate within and across hierarchies. For example, if there were multiple balls of the same color to collect, would participants take the shortest path between them before moving on to a ball of a different color? If there were multiple colors, would they transition from one hierarchical group to another by following the shortest path? These and other questions remain to be explored in future studies using more complex tasks that naturally support (nested) hierarchical structures.} \textcolor{black}{Additionally, future studies could examine more closely the cognitive processes underlying the choice of hierarchical planning—distinguishing, for example, between deliberate decision-making among planning alternatives and more automatic selection driven by simple visual processing and Gestalt principles.} \textcolor{black}{Finally,} investigating how individual differences -- such as cognitive ability or prior experience with spatial tasks -- affect the selection of planning strategies would provide further insight into the underlying mechanisms of decision-making \citep{hegarty2023understanding,weisberg2014variations,boone2018sex,boone2019instructions,santos2018human,krichmar2023importance}.

In conclusion, this study underscores the importance of environmental structure in shaping the strategies individuals adopt when planning and navigating tasks. Notably, we were able to influence participants' preferences by introducing or omitting structure in the environment, with hierarchical planning being favored in structured environments and shortest path strategies in unstructured ones. Additionally, the discrepancy between self-reported preferences and actual behavior suggests that participants may not be fully aware of the implicit strategies they employ, highlighting the complexity of cognitive decision-making processes. Future research should further explore the interaction between environmental cues, cognitive efficiency, and strategy adoption, particularly in more complex and dynamic settings.





\section*{Acknowledgements}

This research received funding from the European Union’s Horizon 2020 Framework Programme for Research and Innovation under the Specific Grant Agreements No. 952215 (TAILOR) to G.P.; the European Research Council under the Grant Agreement No. 820213 (ThinkAhead) to G.P.; the Italian National Recovery and Resilience Plan (NRRP), M4C2, funded by the European Union – NextGenerationEU (Project IR0000011, CUP B51E22000150006, “EBRAINS-Italy”; Project PE0000013, CUP B53C22003630006, "FAIR"; Project PE0000006, CUP J33C22002970002 “MNESYS”) to G.P., and the Ministry of University and Research, PRIN PNRR P20224FESY and PRIN 20229Z7M8N to G.P., and the National Research Council, project iForagers. The funders had no role in study design, data collection and analysis, decision to publish, or preparation of the manuscript. We used a Generative AI model to correct typographical errors and edit language for clarity. 

\section*{Data and Code Availability}
All data and code used to perform the analyses and generate the figures in this study are available at:
\url{https://github.com/DavideNuzzi/Hierarchical-Planning}

\newpage

\bibliographystyle{apalike}
\bibliography{references}

\newpage

\appendix
\setcounter{figure}{0}
\makeatletter 
\renewcommand{\thefigure}{S\@arabic\c@figure}
\makeatother

\setcounter{table}{0}
\makeatletter 
\renewcommand{\thetable}{S\@arabic\c@table}
\makeatother

\section*{Supplementary Materials}

\section{Separate analyses of HN, HS, and HSN trials}
\label{sec:separate}

\begin{figure}
    \centering
    \includegraphics[width=1\linewidth]{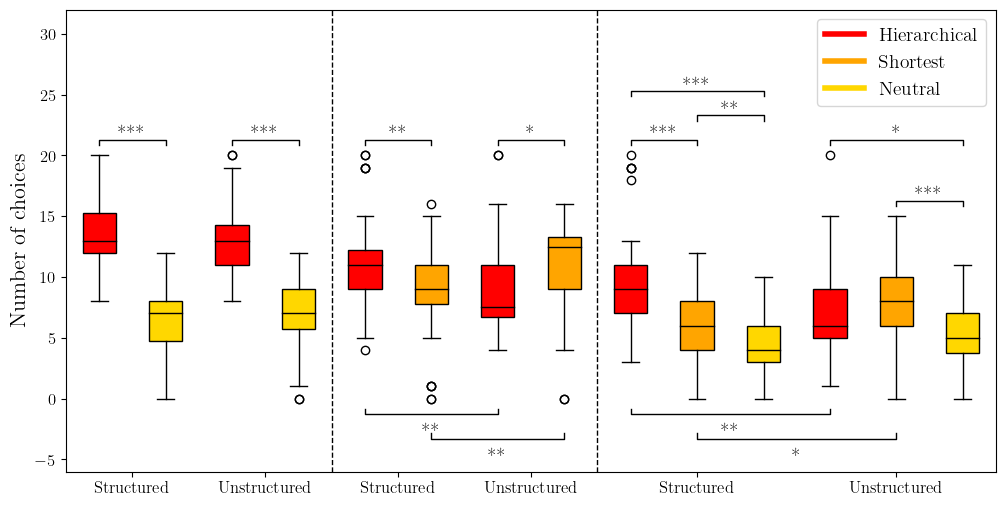}
    \caption{Boxplots showing the number of choices made by participants in structured vs. unstructured environments. The three boxplots display the results for the HN, HS, and HSN trials, respectively. Significance levels were assessed using the Mann-Whitney test, with FDR correction applied. Significant differences between conditions are indicated (*$p < 0.05$, **$p < 0.01$, ***$p < 0.001$). See the main text for further explanation.}
    \label{fig:SeparateAnalysis}
\end{figure}

We performed Mann-Whitney tests with False Discovery Rate (FDR) correction, considering the three trial types (HN, HS, and HSN) separately (Figure \ref{fig:SeparateAnalysis}). 

We first considered the HN trials, where participants selected between hierarchical and neutral plans, in both structured and unstructured blocks. Participants selected significantly more hierarchical plans compared to neutral plans in both structured ($p < 0.001$, \textcolor{black}{$r = 0.95$}) and unstructured blocks ($p < 0.001$, \textcolor{black}{$r = 0.92$}). Consistent with the main text analyses, these results underscore the preference for hierarchical planning when the alternative (neutral) plan has the same path length.

We next considered the HS trials, where participants selected between hierarchical and shortest plans, in both structured and unstructured blocks. Participants selected significantly more hierarchical plans compared to shortest plans in structured blocks ($p < 0.01$, \textcolor{black}{$r = 0.39$}). In contrast, in unstructured blocks, participants selected significantly more shortest plans compared to hierarchical plans ($p < 0.05$, \textcolor{black}{$r = -0.34$}). Furthermore, participants selected more hierarchical plans in structured compared to unstructured blocks ($p < 0.01$, \textcolor{black}{$r = 0.39$}) and more shortest plans in unstructured compared to structured blocks ($p < 0.01$, \textcolor{black}{$r = -0.39$}). Consistent with the main text analyses, these results highlight that hierarchical planning is favored in structured environments, while shortest path planning is more prevalent in unstructured environments. Moreover, the presence of structure enhances the preference for hierarchical planning, whereas its absence increases reliance on shortest path planning.


Finally, we considered the HSN trials, where participants selected between hierarchical, shortest, and neutral plans, in both structured and unstructured blocks. In the structured block, participants selected significantly more hierarchical plans compared to both shortest ($p < 0.001$, \textcolor{black}{$r = 0.53$}) and neutral ($p < 0.001$, \textcolor{black}{$r = 0.8$}) plans, and significantly more shortest plans compared to neutral plans ($p < 0.01$, \textcolor{black}{$r = 0.36$}). In contrast, in the unstructured block, participants selected significantly more hierarchical ($p < 0.05$, \textcolor{black}{$r = 0.31$}) and shortest ($p < 0.001$, \textcolor{black}{$r = 0.47$}) plans compared to neutral plans. Finally, participants selected significantly more hierarchical plans in structured compared to unstructured blocks ($p < 0.01$, \textcolor{black}{$r = 0.4$}) and significantly more shortest plans in unstructured compared to structured blocks ($p < 0.05$, \textcolor{black}{$r = -0.29$}). Consistent with the analyses in the main text, these results show that hierarchical planning prevails over both shortest path and neutral planning in structured blocks. In unstructured blocks, both shortest path and hierarchical planning are favored over neutral planning. The fact that shortest path planning only shows a trend toward being preferred over hierarchical planning \textcolor{black}{($p = 0.18, r = -0.17$)} when the neutral plan was also available may be due to the smaller number of trials in this analysis or increased choice uncertainty. Furthermore, aligning with the main text findings, the presence of environmental structure enhances the preference for hierarchical planning, while its absence increases the preference for shortest path planning.

{\color{black}
\section{Total trial time across conditions}

We analyzed total trial time across the six experimental conditions defined by environment structure (structured vs. unstructured) and trial type (HS, HN, HSN). Total trial time was measured as the duration from trial onset to the collection of the final target sphere and standardized within each participant (z-scored across all their trials) to control for individual differences in baseline speed and variability. In both structured and unstructured environments, HS trials resulted in significantly lower trial times than HN trials (structured: $p < 0.01, r = -0.36$; unstructured: $p < 0.001, r = -0.54$), a pattern that can be explained by the availability of shortest-path strategies in HS trials, which require fewer steps to complete when optimal routes are followed. Both HN trials (structured: $p < 0.001, r = -1$; unstructured: $p < 0.001, r = -1$) and HS trials (structured: $p < 0.001, r = -1$; unstructured: $p < 0.001, r = -1$) were significantly faster than HSN trials, which can be explained by the fact that the latter require collecting one more goal. Furthermore, and more interestingly, we observed significant effects of environmental structure within each trial type: trial times in both HS trials ($p < 0.05, r = -0.29$) and HN trials ($p < 0.001, r = -0.48$) were shorter in structured than in unstructured environments, suggesting that environmental structure supports more efficient task completion. All p-values were computed using the Mann–Whitney U test and corrected for multiple comparisons using the False Discovery Rate (FDR) procedure.

\begin{figure}
    \centering
    \includegraphics[width=1\linewidth]{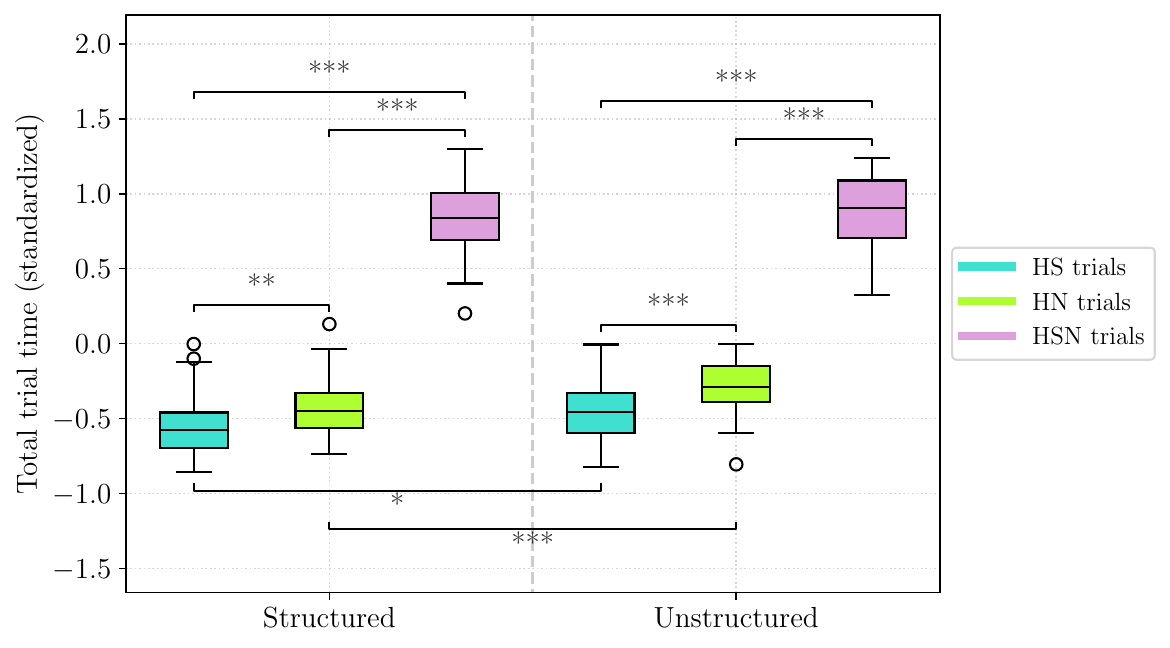}
    \caption{Standardized total trial time across conditions defined by environment structure and trial type. Boxes represent distributions of per-participant z-scored trial durations. Significance levels were assessed using the Mann-Whitney test, with FDR correction applied. Significant differences between conditions are indicated (*$p < 0.05$, **$p < 0.01$, ***$p < 0.001$).}
    \label{fig:TrialTimes}
\end{figure}

}

\end{document}